\documentclass[%
 reprint,
 amsmath,amssymb,
 aps,
]{revtex4-1}

\usepackage{graphicx}
\usepackage{dcolumn}
\usepackage{bm}
\usepackage{float} 

\newcommand*{\affaddr}[1]{#1} 
\newcommand*{\affmark}[1][*]{\textsuperscript{#1}}

\begin{document}

\preprint{APS/123-QED}

\title{Expanding the Reach of Heavy Neutrino Searches at the LHC} 

\author{
Andr\'es Fl\'orez\affmark[2], Kaiwen Gui\affmark[1], Alfredo Gurrola\affmark[1], Carlos Pati\~no\affmark[2], and Diego Restrepo\affmark[3]\\
\affaddr{\affmark[1] Department of Physics and Astronomy, Vanderbilt University, Nashville, TN, 37235, USA}\\
\affaddr{\affmark[2] Physics Department, Universidad de los Andes, Bogot\'a, Colombia}\\
\affaddr{\affmark[3] Department of Physics, Universidad de Antioquia, Medell\'in, Colombia}\\
}

\date{\today}

\begin{abstract}
The observation of neutrino oscillations establishes that neutrinos ($\nu_{\ell}$) have non-zero mass and provides one of the more compelling arguments for physics beyond the standard model (SM) of particle physics. We present a feasibility study to search for hypothetical Majorana neutrinos ($N_{\ell}$) with TeV scale masses, predicted by extensions of the SM to explain the small but non-zero $\nu_{\ell}$ mass, using vector boson fusion (VBF) processes at the 13 TeV LHC. In the context of the minimal Type-I seesaw mechanism (mTISM), the VBF $\ell N_{\ell}$ production cross-section surpasses that of the Drell-Yan process at approximately $m_{N_{\ell}} = 1.4$ TeV. We consider $\mu N_{\mu}$ and $\tau N_{\tau}$ production through VBF processes (e.g. $qq'\to \tau N_{\tau}qq'$), with subsequent $N_{\mu}$ and $N_{\tau}$ decays to $\mu jj$ and $\tau jj$, as benchmark cases to show the effectiveness of the VBF topology for $N_{\ell}$ seaches at the 13 TeV LHC. The requirement of a dilepton pair combined with four jets, two of which are identified as VBF jets with large separation in pseudorapidity and a TeV scale dijet mass, is effective at reducing the SM background. This criteria may provide expected exclusion bounds, at 95\% confidence level, of $m_{N_{\ell}} < 1.7$ ($2.4$) TeV, assuming $100$ (1000) fb$^{-1}$ of 13 TeV data from the LHC and mixing $|V_{\ell N_{\ell}}|^{2} = 1$. The use of the VBF topology to search for $m_{N_{\ell}}$ increases the discovery reach at the LHC, with expected significances greater than 5$\sigma$ (3$\sigma$) for $N_{\ell}$ masses up to 1.7 (2.05) TeV using 1000 fb$^{-1}$ of 13 TeV data from the LHC.

\end{abstract}

\pacs{Valid PACS appear here}
\maketitle

\section{\label{sec:level1}Introduction}

The discovery of a Higgs boson~\cite{Aad20121,Chatrchyan201230} at the Large Hadron Collider (LHC) has addressed the last missing piece of the standard model (SM) of particle physics. However, the SM remains an incomplete theory. One of the open questions it fails to address is the non-zero mass of the three generations of neutrinos, which is implied by the observation of neutrino oscillations~\cite{Fukuda:1998mi,Ahmad:2001an,Ahmad:2002jz}. It has been suggested that because neutrinos can be their own anti-particles (Majorana fermions), the non-zero mass of light neutrinos $\nu_{\ell}$ could be generated by a see-saw mechanism~\cite{Lindner:2001hr,Minkowski:1977sc,Mohapatra:1979ia}, which would imply the existence of yet unobserved heavier Majorana neutrino states with TeV scale masses. For example, in the left-right symmetric model (LRSM), originally introduced to explain the non-conservation of parity in weak interactions within the SM, the introduction of a SU(2)$_{R}$ group, the right-handed analogue of the SM SU(2)$_{L}$ group, produces three heavy right-handed neutrino states $N_{\ell}$ ($\ell = e,\mu,\tau$) and three gauge bosons, $V_{R} = \{ W_{R}^{\pm}, Z' \}$.

The CMS~\cite{Chatrchyan:2008aa} and ATLAS~\cite{Aad:2008zzm} experiments at the CERN LHC have a strong physics program to search for heavy right-handed neutrinos. One often used benchmark model in those searches is the LRSM. Within this context, the CMS and ATLAS searches assume that $N_{\ell}$ is lighter than $V_{R}$: $m_{N_{\ell}} < m_{W_{R}^{\pm}}$ and $m_{N_{\ell}} < 0.5m_{Z'}$. Under this assumption, the dominant $N_{\ell}$ production mechanism at the LHC is via resonant $W_{R}^{\pm}$ or $Z'$ production from Drell-Yan (DY) processes of order $\alpha_{EW}^{2}$: $qq' \to W_{R}^{\pm} \to \ell N_{\ell}$ or $q\bar{q} \to Z' \to N_{\ell} N_{\ell}$. The strategy pursued in those analyses is to exploit the high mass scale of $V_{R}$ and target the $N_{\ell}$ decay to a lepton and two jets (through a virtual $W_{R}$), $N_{\ell}\to \ell W_{R}^{*} \to \ell j j$, by selecting events containing two high-$p_{T}$ leptons (opposite-sign or like-sign charge) and two jets that are central in the detector (i.e. pseudorapidity range $|\eta| < 3.0$). Therefore, dilepton triggers can be used to select signal events with high efficiency. Furthermore, because DY-like production of resonant $W_{R}^{\pm}$ or $Z'$ is dominant if $m_{N_{\ell}} < m_{W_{R}^{\pm}}$ or $m_{N_{\ell}} < 0.5m_{Z'}$, the invariant mass distribution of the system consisting of two high-$p_{T}$ leptons and jets, $m_{\ell \ell \Sigma{j}}$, produces a "bump" in signal events, at $m_{\ell \ell \Sigma{j}} \approx m_{V_{R}}$, which can be utilised to discriminate against the smooth and steeply falling SM background distribution. Results of those searches in proton-proton collisions at $\sqrt{s}=7$, 8, and 13 TeV exclude $N_{\ell}$ masses below 1.5 (0.8 TeV)~\cite{ATLAS:2012ak,Khachatryan:2014dka,Khachatryan:2016jqo}, assuming $m_{N_{\ell}}$ is 0.5 (0.4) times the mass of the $W_{R}$ ($Z'$) boson. However, the sensitivity to $N_{\ell}$ is dependent on the mass of the $W_{R}^{\pm}$ or $Z'$ bosons. For example, no bounds on $m_{N_{\ell}}$ exist for $m_{W_{R}^{\pm}}$ ($m_{Z'}$) $\sim m_{N_{\ell}}$ ($2 m_{N_{\ell}}$). Additionally, if the $W_{R}^{\pm}$ and $Z'$ bosons are too heavy to provide large enough cross-section, regardless of $m_{N_{\ell}}$, another technique must be devised to probe $N_{\ell}$. For this reason, the CMS and ATLAS experiments also perform more general and less model-dependent searches by considering the simplest extension to the SM which can explain the small but non-zero mass of $\nu_{\ell}$, the so-called minimal Type-I seesaw mechanism (mTISM). In this model, the only additional degrees of freedom beyond the SM are the heavy Majorana neutrinos. Therefore, the $N_{\ell}$ is produced through an off-mass-shell $W$ boson, $qq'\to W^{*} \to \ell N_{\ell}$. Although the $\ell \ell jj$ final state particles are similar to the case of the LRSM, the kinematics are different. In the mTISM final state, because the first lepton is produced by an off-mass-shell  $W$ boson, $p_{T}(\ell)$ is significantly smaller on average than in the LRSM final state. Also, because the jets are produced by an on-mass-shell $W$ boson ($N_{\ell} \to \ell W \to \ell jj$), the dijet mass is consistent with $m_{W}$. Results of the mTISM searches depend on the mixing between $N_{\ell}$ and $\nu_{\ell}$, $|V_{\ell N_{\ell}}|^{2}$, and exclude $m_{N_{\ell}} < 500$ ($200$) GeV for $|V_{\ell N_{\ell}}|^{2} = 1$ ($10^{-2}$).

\begin{figure}
 \begin{center} 
 \includegraphics[width=0.5\textwidth, height=0.3\textheight]{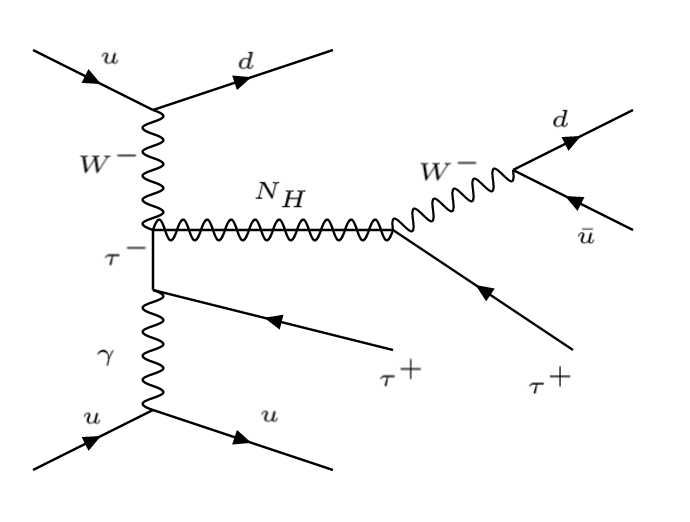}
 \end{center}
 \caption{Feynman diagram depicting pure electroweak production of a $N_{\ell}$ particle through VBF.}
 \label{fig:feyn}
 \end{figure}

The focus of this paper is to explore production of $N_{\ell}$ via vector boson fusion (VBF) processes and the effectiveness of the VBF topology for $N_{\ell}$ seaches at the 13 TeV LHC. A search for $N_{\ell}$ using the VBF topology has not been performed before at a collider, but may present an important avenue for discovery. The VBF topology is characterized by two high $p_{T}$ forward jets ($j_{f}$), with large pseudorapidity gap, located in opposite hemispheres of the detector, and TeV scale dijet invariant masses. The tagging of events produced though VBF processes has been proposed by some of the present authors as an effective experimental tool for dark matter (DM) and electroweak supersymmetry (SUSY) searches at the LHC~\cite{VBF1,DMmodels2,VBFSlepton,VBFStop,VBFSbottom}, as well as searches for a TeV scale neutral gauge boson $Z'$~\cite{Andres:2016xbe}. A comparison with current results from CMS and ATLAS is performed in the most model-independent setting by using the mTISM model as the benchmark scenario. To highlight the usefulness of the VBF topology in $N_{\ell}$ searches with large QCD backgrounds as well as those with cleaner signatures, we consider $\mu N_{\mu} j_{f}j_{f}$ and $\tau N_{\tau} j_{f}j_{f}$ production through VBF processes with subsequent $N_{\mu}$ and $N_{\tau}$ decays to $\mu jj$ and $\tau jj$, respectively. In the VBF $N_{\tau}$ study, we focus on the final state where both $\tau$ leptons decay to hadrons ($\tau_{h}$) since it provides the largest branching fraction (42\%) compared to final states with semi-leptonic decays of $\tau$ leptons. Therefore, the final states considered in this paper are $\mu\mu jj j_{f}j_{f}$ and $\tau_{h}\tau_{h} jj j_{f}j_{f}$, where $j$ refers to a jet not tagged as a VBF jet. Figure~\ref{fig:feyn} shows an example Feynman diagram for the production mechanism of $N_{\tau}$, in particular $W\gamma$ fusion in t-channel diagrams containing an off-mass shell $W$ boson. The $N_{\tau}$ subsequently decays to a $\tau$ lepton and two jets.

\section{Samples and simulation}

\begin{figure}[H]
 \begin{center} 
 \includegraphics[width=0.5\textwidth, height=0.3\textheight]{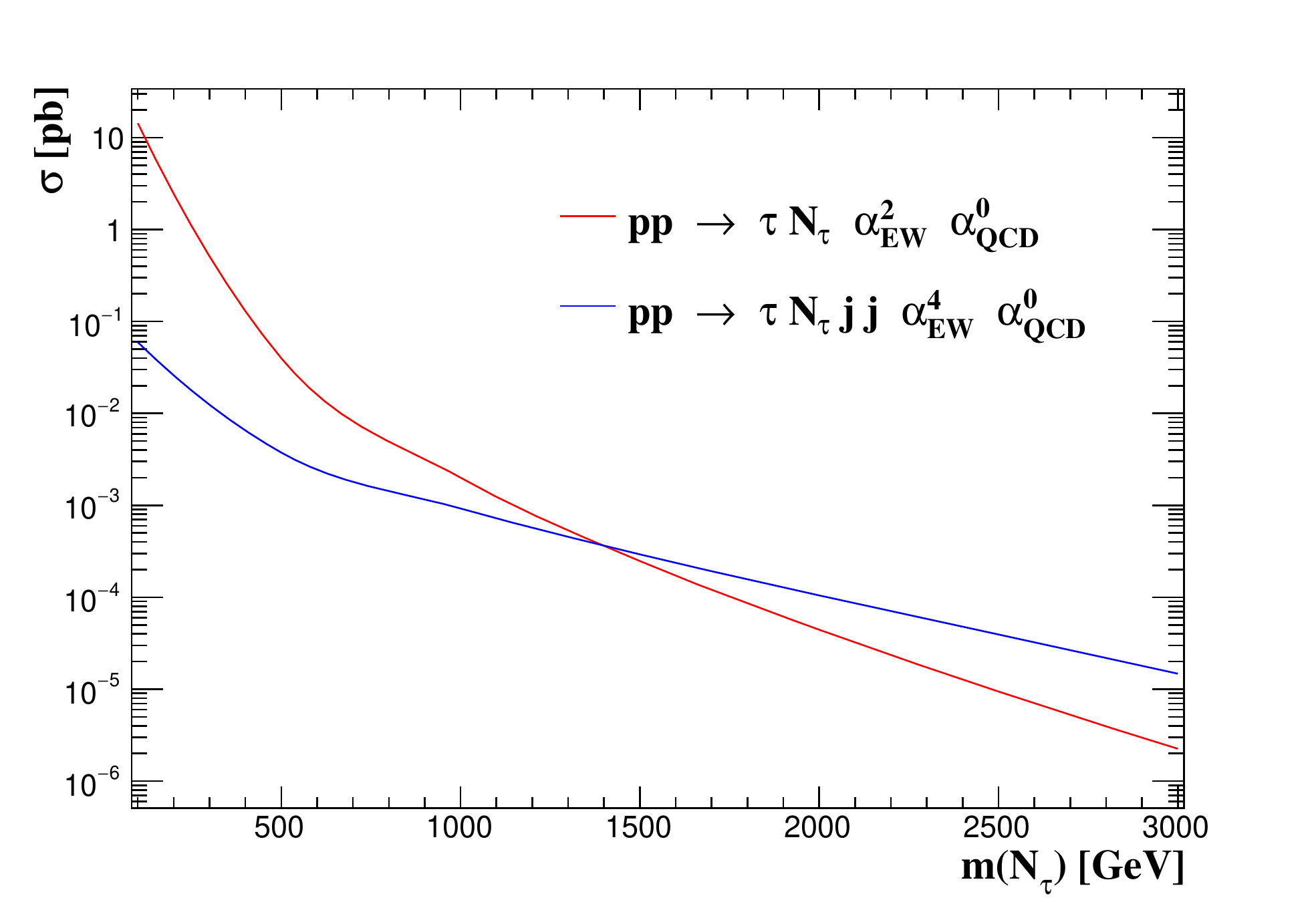}
 \end{center}
 \caption{$N_{\ell}$ production cross-section as a function of mass. The VBF $N_{\ell}$ production cross-section exceeds the production via the DY process at approximately $m_{N_{\ell}} = 1.4$ TeV. }
 \label{fig:XSections}
 \end{figure}

The SM background and $N_{\ell}$ signal event samples were generated with MadGraph (v2.2.3) \cite{MADGRAPH}. We considered two sets of signal samples: (i) $pp \to \mu N_{\mu} j_{f} j_{f}$, and (ii) $pp \to \tau N_{\tau} j_{f} j_{f}$ via pure electroweak VBF processes of order $\alpha_{EW}^{4}$. The $N_{\ell}$ masses considered range from 0.5 TeV to 3 TeV in steps of 0.25 TeV. 
At the MadGraph level, leptons were required to have a $p_{T} (\ell) > 10$ GeV and $|\eta (\ell)| < 2.5$, while jets were required to have a minimum $p_{T}> 20$ GeV and $|\eta| < 5.0$.  Figure~\ref{fig:XSections} shows a comparison of the $pp \to \tau N_{\tau}$ (DY-like) and $pp \to \tau N_{\tau} j_{f}j_{f}$ (VBF) production cross-sections as a function of $m_{N_{\tau}}$. At the 13 TeV LHC, the VBF $\ell N_{\ell}$ production cross-section surpasses that of the DY process at approximately $m_{N_{\ell}} = 1.4$ TeV. We note the $pp \to \mu N_{\mu} j_{f}j_{f}$ production cross-section is similar to the $pp \to \tau N_{\tau} j_{f}j_{f}$ cross-section shown in Figure~\ref{fig:XSections} as long as the mixing parameters are equal.


The dominant sources of background in these studies are the production of top quark pairs with associated jets from initial state radiation (ISR) processes ($t\bar{t}$), $Z/\gamma*\to{\ell}{\ell}$ with associated jets mostly from ISR ($Z$+jets), and events with a $W$ boson and ISR jets ($W$+jets). Samples of events from the production of pairs of vector bosons ($VV$) were also considered as a potential source of background, but were found to be negligible after the signal selection criteria outlined in subsequent sections. For this reason they are not included in the figures. Background from $t\bar{t}$ events is characterised by two b quark jets from the decays of the top quarks, two real prompt isolated leptons from the decay of a $W$ boson ($t\to b W \to b (\ell \nu_{\ell})$), and two additional jets from initial state radiation. 
Although $Z$+jets events can be produced through VBF processes, the requirement of four jets in the final state suppresses the VBF $Z$ contribution and makes $Z$ production with associated ISR jets the dominant contribution to the $Z$+jets background. Since the $Z/\gamma^{*}$ bosons can subsequently decay to real leptons, the topology of this background includes two real leptons and four additional ISR jets, two of which are in the forward parts of the detector. In the case of $W$+jets events, which is only relevant for the $\tau_{h}\tau_{h} jj j_{f}j_{f}$ final state, a real prompt lepton is obtained from the decay of the $W$ boson and the second $\tau_{h}$ results from the misidentification of a jet as a $\tau_{h}$. 

PYTHIA (v6.416) \cite{Sjostrand:2006za} was used for the hadronization process of the signal and background samples. The Delphes (v3.3.2) \cite{deFavereau:2013fsa} framework was used to simulate detector effects using the CMS configuration. The $t\bar{t}$ background sample was generated with up to two associated jets, while the $Z$+jets and $W$+jets samples were generated with up to four associated jets, inclusive in $\alpha_{EW}$ and $\alpha_{QCD}$. The MLM algorithm \cite{MLM} was used for jet matching and jet merging, which optimizes two variables (xqcut and qcut) related to the jet definition. The xqcut variable defines the minimal distance between partons at MadGraph level. The qcut variable defines the minimum energy spread for a clustered jet in PYTHIA. In order to determine appropriate xqcut and qcut values, the distribution of the differential jet rate was required to smoothly transition between events with $N$ and $N+1$ jets. The jet matching and merging studies resulted in an optimized xqcut and qcut of 30. 

\section{Event selection criteria}

\begin{figure}
 \begin{center} 
 \includegraphics[width=0.45\textwidth, height=0.3\textheight]{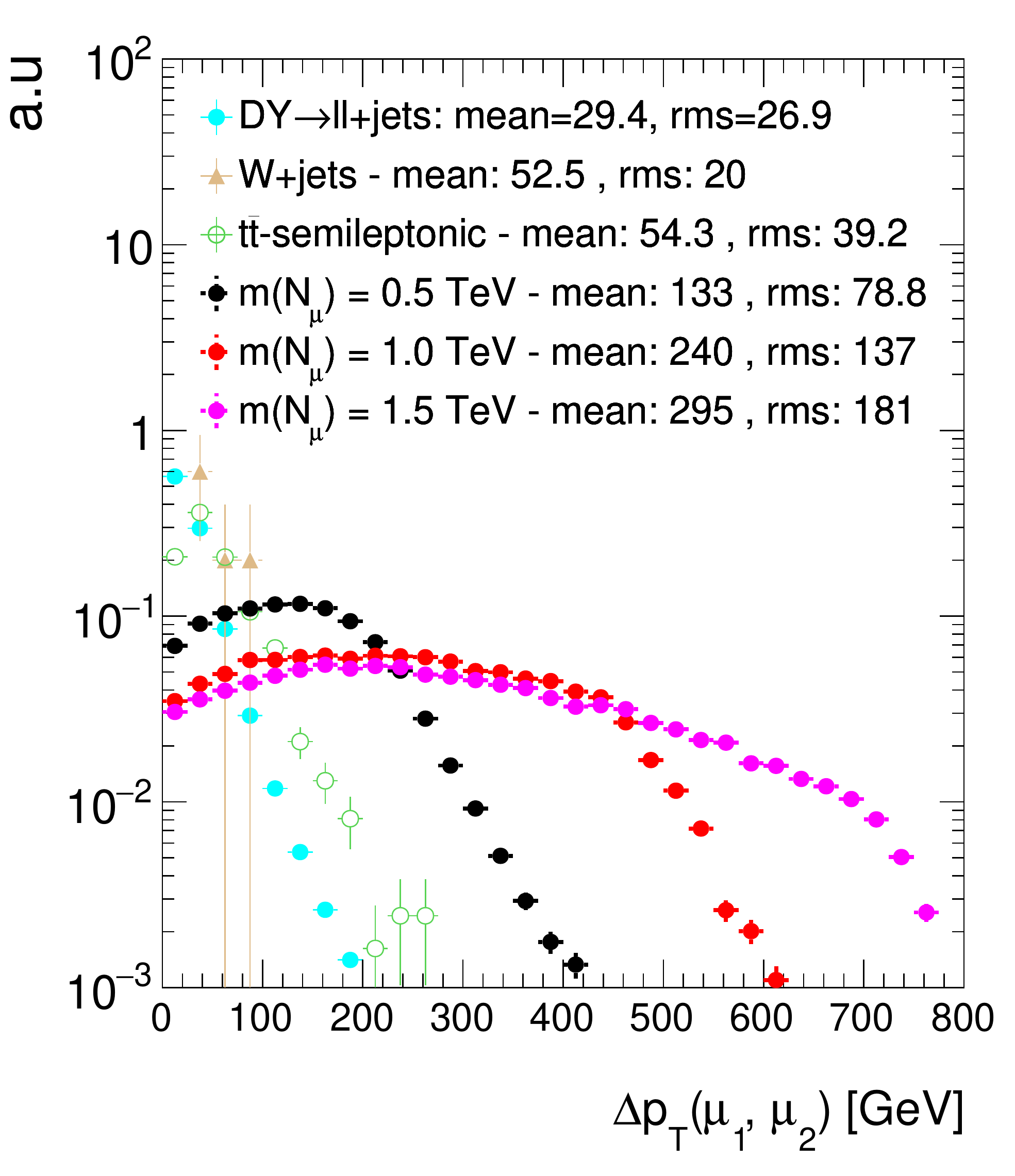}
 \end{center}
 \caption{$\Delta p_{T}$ distributions, normalized to unity, for the $\mu\mu j j j_{f} j_{f}$ final state. The distributions are obtained after requiring at least two muons with $p_{T} > 10$ GeV and $|\eta| < 2.4$.}
 \label{fig:kinDistriNorm1}
 \end{figure}

The event selection criteria used in these studies are divided in two parts, referred to as central selections and VBF selections. The central selections include requirements on the transverse momentum of the leptons and jets ($p_{T} (\ell/j) > X$), the geometric acceptance pseudorapidity requirements ($|\eta (\ell/j)| < Y$), the absolute value of the scalar difference in $p_{T}$ between the two lepton candidates ($\Delta p_{T} = |p^{\ell 1}_{T} - p^{\ell 2}_{T}|$), the product of the electric charges of the two leptons ($Q(\ell 1) \times Q(\ell 2)$), and a veto on the number of jet candidates identified as b-quarks ($N_{b} = 0$). Since our focus is high-mass $N_{\ell}$, signal events are characterized by one high-$p_{T}$ lepton ($p_{T} ({\ell}) \sim \frac{1}{3}m_{N_{\ell}}$). Therefore, the highest-$p_{T}$ lepton in the event is required to have $p_{T} > 50$ GeV, which helps to drastically reduce events from $Z$+jets and $W$+jets processes (e.g. $p_{T}(\tau_{h}) \sim m_{W}/4 = 20$ GeV in $W$+jets events). The $p_{T}$ cut on the second lepton is driven by the experimental constraints of the CMS and ATLAS reconstruction algorithms. The sub-leading muon must have $p_{T} > 10$ GeV, while the sub-leading $\tau_{h}$ must have $p_{T} > 20$ GeV. Events are required to have at least four jets, two of which are considered central jets. The pseudorapidity is constrained to $|\eta| < 2.4$ $(2.1)$ for muons ($\tau_{h}$) and $|\eta| < 2.4$ for central jets, which allows these objects to be within the tracker coverage of the detector, thus improving reconstruction efficiency. In background events, the leptons mainly come from decays of $W$ bosons (e.g. $t\bar{t}\to bb WW \to bb \ell \nu_{\ell} \ell \nu_{\ell}$) and $Z/\gamma^{*}$ bosons ($Z\to \ell \ell$), and therefore both leptons have a similar $p_{T}$ on average, i.e. $p_{T}(\ell_{1}) \approx p_{T}(\ell_{2})$. On the other hand, because the first lepton in signal events is produced via non-resonant t-channel $W\gamma$ fusion diagrams (making it low-$p_{T}$) and the second lepton is produced by the decay of a heavy on-mass shell $N_{\ell}$ (making it high-$p_{T}$), the $p_{T}$ values of the two leptons are largely asymmetric. This motivates a requirement on the absolute value of the scalar difference in $p_{T}$ between the two leptons. Figure~\ref{fig:kinDistriNorm1} shows the $\Delta p_{T} = |p^{\ell 1}_{T} - p^{\ell 2}_{T}|$ distribution for the $\mu\mu j jj j_{f} j_{f}$ final state, normalized to unity, for two signal benchmark points and the main backgrounds. A $\Delta p_{T} = |p^{\ell 1}_{T} - p^{\ell 2}_{T}| > 50$ GeV criterion helps to suppress approximately 80\% of the SM background, while maintaining about a 90\% efficiency for signal events. A same-sign charge requirement between the two leptons, $Q(\ell 1) \times Q(\ell 2) > 0$, reduces the $Z$+jets background by two orders of magnitude and is over 95\% efficient for signal. Finally, requiring events with zero jets tagged as b-quarks suppresses 80\% of the $t\bar{t}$ background, while being $\sim$ 90\% efficient for signal events.

\begin{figure}
 \begin{center} 
 \includegraphics[width=0.45\textwidth, height=0.3\textheight]{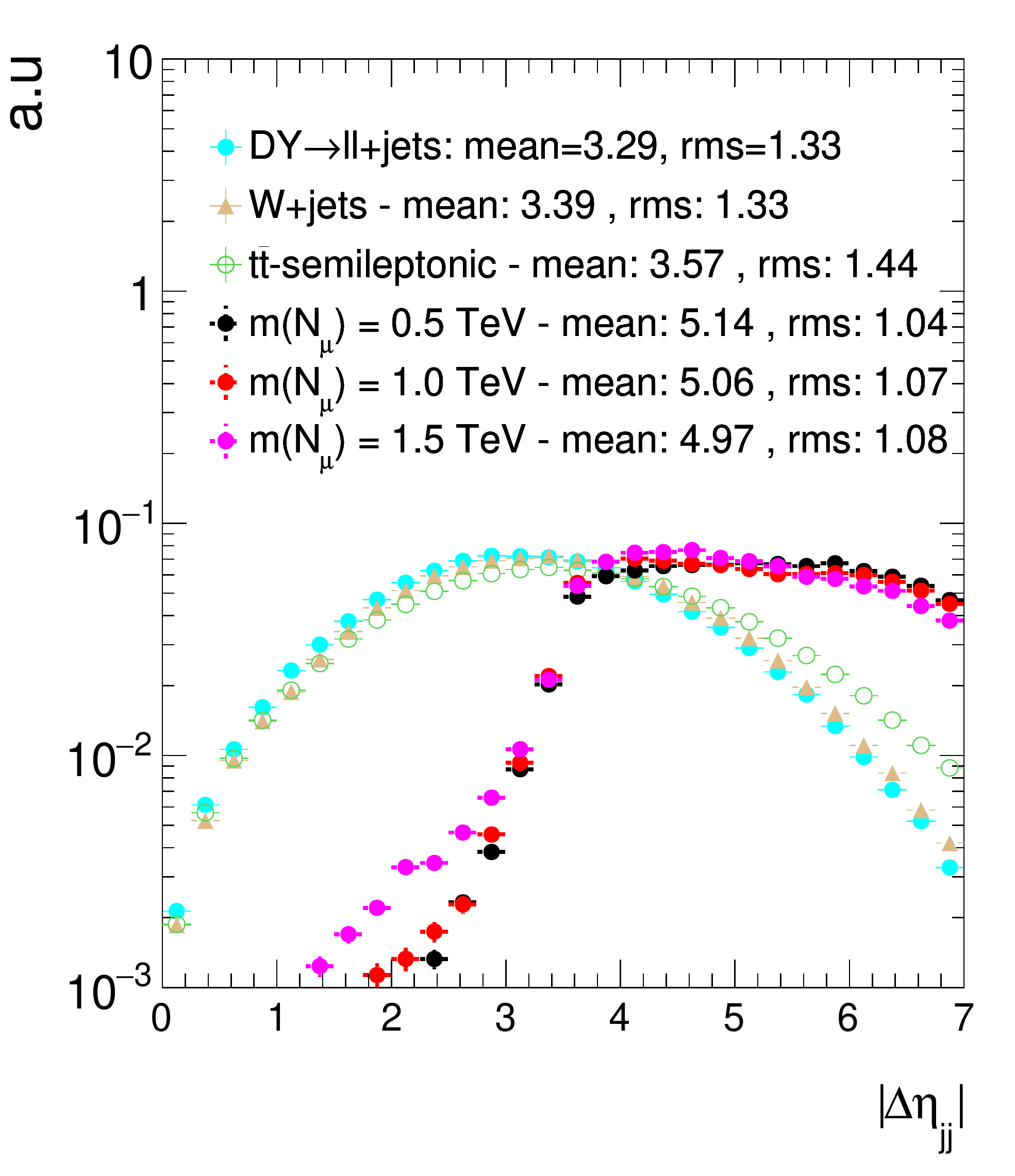}
 \end{center}
 \caption{Dijet $|\Delta \eta|$ distributions, normalized to unity, for the $\tau_{h}\tau_{h} j_{f} j_{f}$ final state. The distributions are obtained after requiring at least two $\tau_{h}$ candidates with $p_{T} > 20$ GeV and $|\eta| < 2.1$.}
 \label{fig:kinDistriNorm2}
 \end{figure}

\begin{figure}
 \begin{center} 
 \includegraphics[width=0.45\textwidth, height=0.3\textheight]{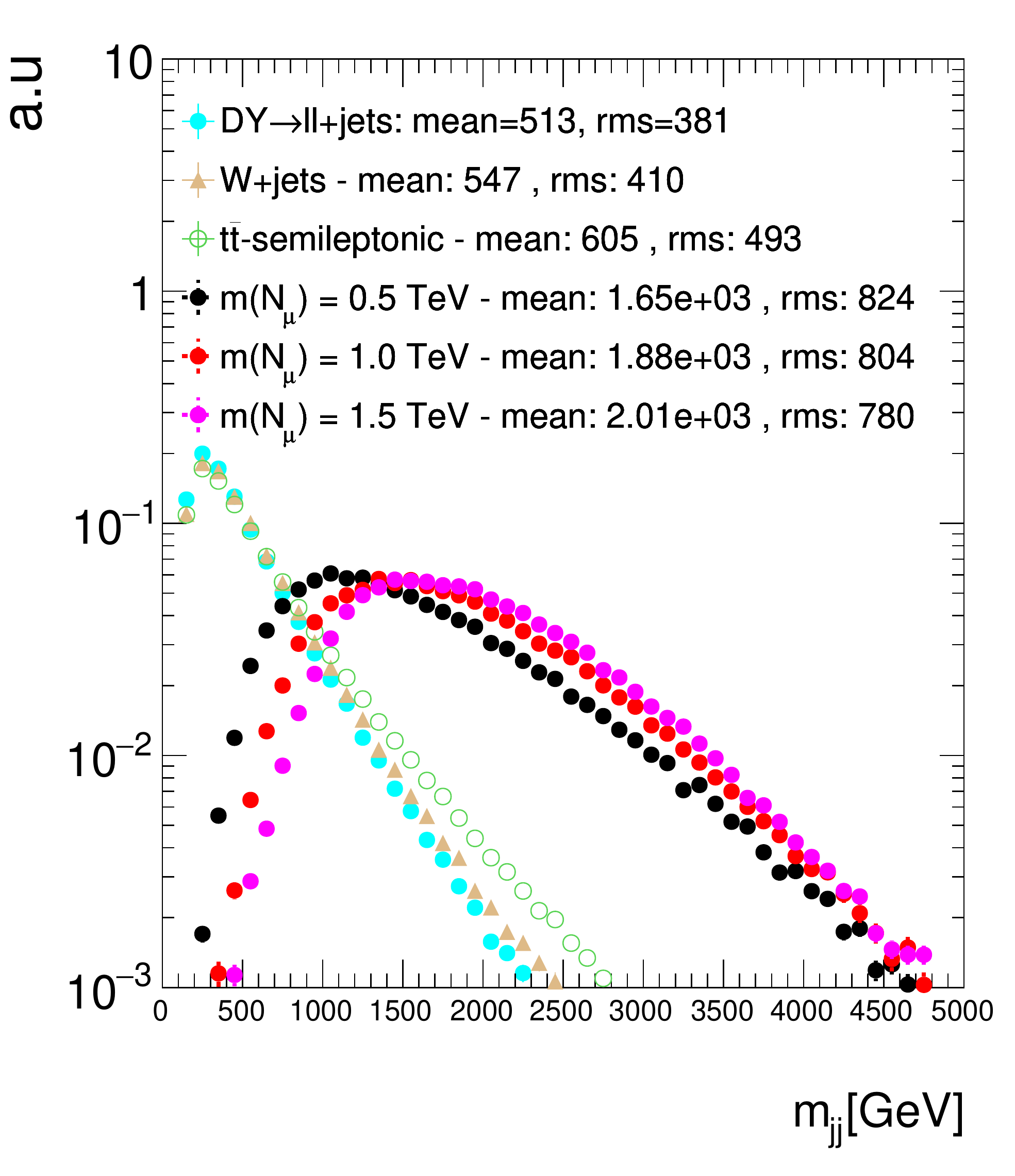}
 \end{center}
 \caption{Dijet mass distributions, normalized to unity, for the $\tau_{h}\tau_{h} j_{f} j_{f}$ final state. The distributions are obtained after requiring at least two $\tau_{h}$ candidates with $p_{T} > 20$ GeV and $|\eta| < 2.1$, and two jets with 
 $p_{T} > 30$ GeV, $|\eta| < 5.0$, $|\Delta \eta_{jj}| > 4.2$ and $m_{jj} > 500$ GeV.}
 \label{fig:kinDistriNorm3}
 \end{figure}

The distinctive signature of VBF processes is the presence of two high-$p_{T}$ jets, with a large pseudorapidity gap ($|\Delta \eta|$), and located in opposite hemispheres of the detector. The invariant mass of the $j_{f} j_{f}$ pair, especially for signal, is expected to be broad and fall at TeV scale values. In addition to requiring at least four jets in the event topology, the VBF selection criteria requires two of those four jets have $p_{T} > 30$ GeV, $|\eta| < 5.0$, $|\Delta \eta_{j_{f}j_{f}}| > 4.2$, $\eta_{j_{f,1}} \cdot \eta_{j_{f,2}} < 0$, and $m_{j_{f}j_{f}} > 750$ GeV. Figure~\ref{fig:kinDistriNorm2} shows the $|\Delta \eta_{j_{f}j_{f}}|$ distribution in the $\tau_{h} \tau_{h} j j j_{f} j_{f}$ channel, after requiring the central selections described above. Only the dijet pair with the largest $m_{j_{f}j_{f}}$ is chosen to populate the distributions in Figure~\ref{fig:kinDistriNorm2}. Figure~\ref{fig:kinDistriNorm3} shows the invariant dijet mass distribution, with the same criteria as in Figure~\ref{fig:kinDistriNorm2}. 
The VBF selection criteria is a powerful tool to reduce the contribution from $t\bar{t}$ (rejection power of $10^{2}$) and $Z/W$+jets (rejection factors of $10^{3}$-$10^{4}$). It is further noted that the QCD multijet background is not shown in those figures because it is found to be negligible with only the central selections (i.e. negligible yield after requiring six objects: two isolated leptons and four jets). However, VBF is also important to ensure the suppression of the QCD multijet background ($10^{4}$ suppression factor), which are typically relevant in final states with $\tau_{h}$ candidates due to the high jet-$\tau_{h}$ misidentification rate~\cite{VBF2,CMSVBFDM}. Table~\ref{tab:selections} summaries the selections used in the $\mu\mu j j j_{f}j_{f}$ and $\tau_{h} \tau_{h} j j j_{f}j_{f}$ final states.

\begin{table}
\begin{center}
\caption {Event selection criteria for the $\mu\mu / \tau\tau$ $+$ $jjj_{f}j_{f}$ channels.}
\label{tab:selections}
\begin{tabular}{ l  c c}\hline\hline
Criterion & $\tau_{h}\tau_{h} jjj_{f}j_{f}$ &  $\mu\mu jjj_{f}j_{f}$\\
 \hline
  \multicolumn{3}{ c }{{\bf Central Selections}} \\
   \hline
  $|\eta(\tau_{h}/\mu)|$ &  $< 2.1$ &  $< 2.4$\\  
  $p^{lead}_{T}(\tau_{h}/\mu)$ & $> 50$ GeV & $>  50$ GeV\\
  $p^{slead}_{T}(\tau_{h}/\mu)$ & $> 20$ GeV & $>  10$ GeV\\
  $N(\tau_{h}/\mu)$ & $\ge 2$ & $\ge 2$ \\
  $\Delta R(\tau_{h1}/\mu_{1},\tau_{h2}/\mu_{2})$ & $> 0.3$ & $> 0.3$ \\ 
   $|\Delta p_{T} (\tau_{h1}/\mu_{1}, \tau_{h2}/\mu_{2})|$ & $> 50$ GeV & $> 50$ GeV \\
   $Q(\tau_{h1}/\mu_{1}) \times Q(\tau_{h2}/\mu_{2}) $ & $> 0$ & $> 0$ \\
   $N_{b-jets}$ & $= 0$ & $= 0$ \\
   $p^{central}_{T}(j)$ & $30$ GeV & $30$ GeV \\
   $|\eta^{central}(j)|$ & $ < 2.4$ & $< 2.4$ \\
   $N_{central}(j)$ & $=2$ & $=2$ \\
   $\Delta R(\tau_{h}/\mu, j)$ & $> 0.4$ & $> 0.4$ \\ 
   \hline
   \multicolumn{3}{ c }{{\bf VBF Selections}} \\
   \hline
  $p^{lead}_{T}(j_{f})$ & 30 GeV & 30 GeV \\
   $|\eta^{lead} (j_{f})|$ & $< 5.0$  & $< 5.0$\\
   $p^{sub-lead}_{T}(j_{f})$ & 30 GeV & 30 GeV \\
   $|\eta^{sub-lead} (j_{f})|$ & $< 5.0$  & $< 5.0$\\
   $\Delta R(\tau_{h}/\mu, j_{f})$ & $> 0.4$ & $> 0.4$ \\
   $\eta(j_{f,1})\cdot \eta(j_{f,2})$ & $< 0$ & $< 0$ \\
   $|\Delta \eta (j_{f,1}, j_{f,2})| $ & $> 4.2$ & $> 4.2$\\
   $m_{j_{f}j_{f}}$ & $> 750$ GeV & $> 750$ GeV \\
   \hline\hline
 \end{tabular}
\end{center}
\end{table}

\section{Results}

\begin{figure}
 \begin{center} 
 \includegraphics[width=0.5\textwidth, height=0.35\textheight]{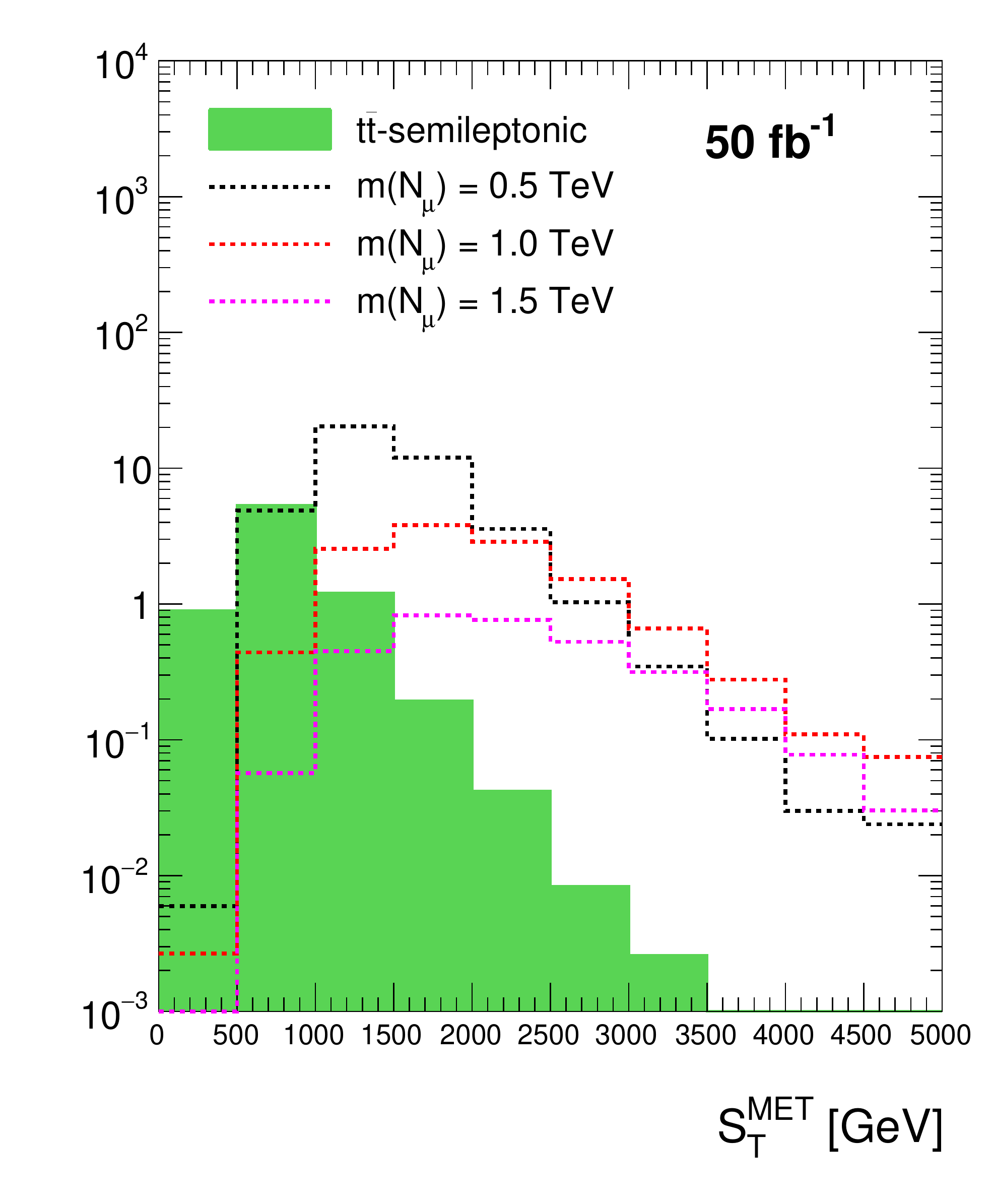}
 \end{center}
 \caption{$S^{MET}_{T}$ distribution in the $\mu\mu j j j_{f} j_{f}$ final state, for the main backgrounds and two chosen signal benchmark points, after applying the final event selection criteria. The distributions are normalized to an integrated luminosity of $50$ fb$^{-1}$.}
 \label{fig:STmetMuon}
 \end{figure}

\begin{figure}
 \begin{center} 
 \includegraphics[width=0.5\textwidth, height=0.35\textheight]{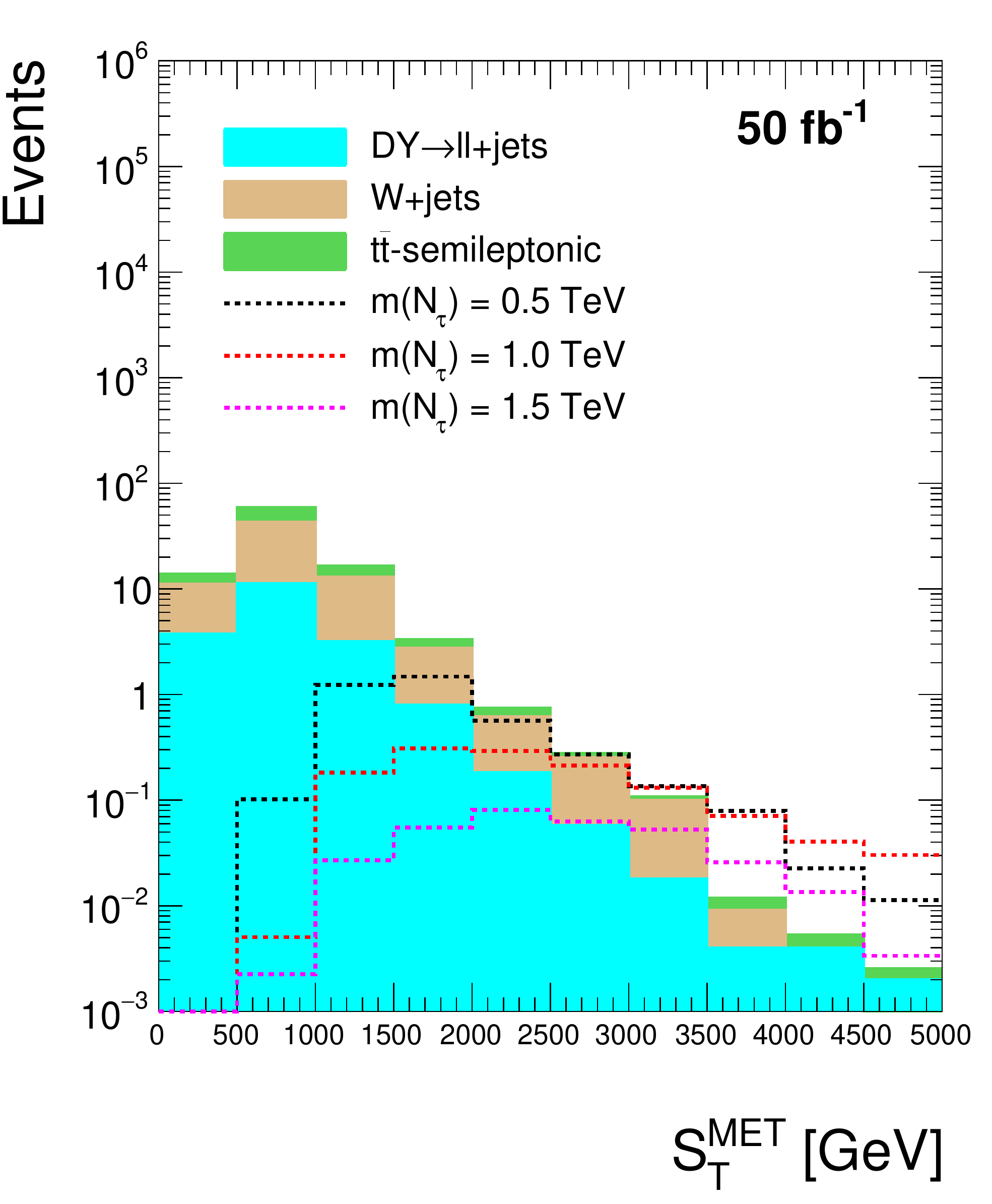}
 \end{center}
 \caption{$S^{MET}_{T}$ distribution in the $\tau_{h}\tau_{h} j j j_{f} j_{f}$ final state, for the main backgrounds and two chosen signal benchmark points, after applying the final event selection criteria. The distributions normalized to an integrated luminosity of $50$ fb$^{-1}$.}
 \label{fig:STmetTau}
 \end{figure}

The expected experimental sensitivity of each final state is determined using a binned-likelihood approach (i.e. a shape based analysis instead of a cut and count approach) following the test statistic based on the profile likelihood ratio, using the ROOTFit \cite{ROOTFit} toolkit. The sensitivity is determined using the ``fit variable" that gives the best signal significance $z$. The signal significance is determined by first calculating a local p-value, defined as the the probability under a background only hypothesis to obtain a value of the test statistic as large as that obtained with a signal plus background hypothesis, and then extracting the value at which the integral of a Gaussian between $z$ and $\infty$ results in a value equal to the local p-value. Systematic uncertainties are incorporated via nuisance parameters following the frequentist approach. The dominant systematic uncertainties considered - $\tau_{h}$ identification (6\%), VBF selection efficiency (20\%), and the uncertainty due to the variations in the yields and shapes arising from the choice of parton distribution function (15\%) - are based on available ATLAS and CMS results using $\tau_{h}$ candidates, muons, and/or the VBF topology~\cite{CMSTauID,VBF2,CMSVBFDM,ATLASZprimeMuMuPAS,CMSZprimeMuMuPAS,ATLASZprimeTauTau,CMSZprimeTauTauPAS}. Based on these considerations, a total systematic uncertainty of 25\% is applied on the signal and background yields. Several kinematic and topological variables, such as the missing transverse momentum $p^{miss}_{T}$, $m_{j_{f}j_{f}}$, $H_{T}$, $S_{T}$ and $S^{MET}_{T}$, were considered as possible fit variables. The $H_{T}$ variable is defined as the scalar sum of the $p_{T}$ of all the jets in the event. The $S_{T}$ variable is defined as the scalar sum of the $H_{T}$ and the $p_{T}$ of the two selected lepton candidates ($\tau_{h} \tau_{h}$ or $\mu \mu$ depending on the channel under consideration). The $S^{MET}_{T}$ variable represents the scalar sum of the $S_{T}$ and the $p^{miss}_{T}$ variables. Using $z$ as the figure of merit, the $S^{MET}_{T}$ distribution provides the best expected sensitivity in both final states. Figures~\ref{fig:STmetMuon} and~\ref{fig:STmetTau} show the $S^{MET}_{T}$ distributions, after all of the event selection criteria outlined in Table~\ref{tab:selections} and normalized to an integrated luminosity of 50 fb$^{-1}$, for the $\mu \mu j j j_{f} j_{f}$ and the $\tau_{h} \tau_{h} j j j_{f} j_{f}$ search channels, respectively.

\begin{figure}
 \begin{center} 
 \includegraphics[width=0.5\textwidth, height=0.35\textheight]{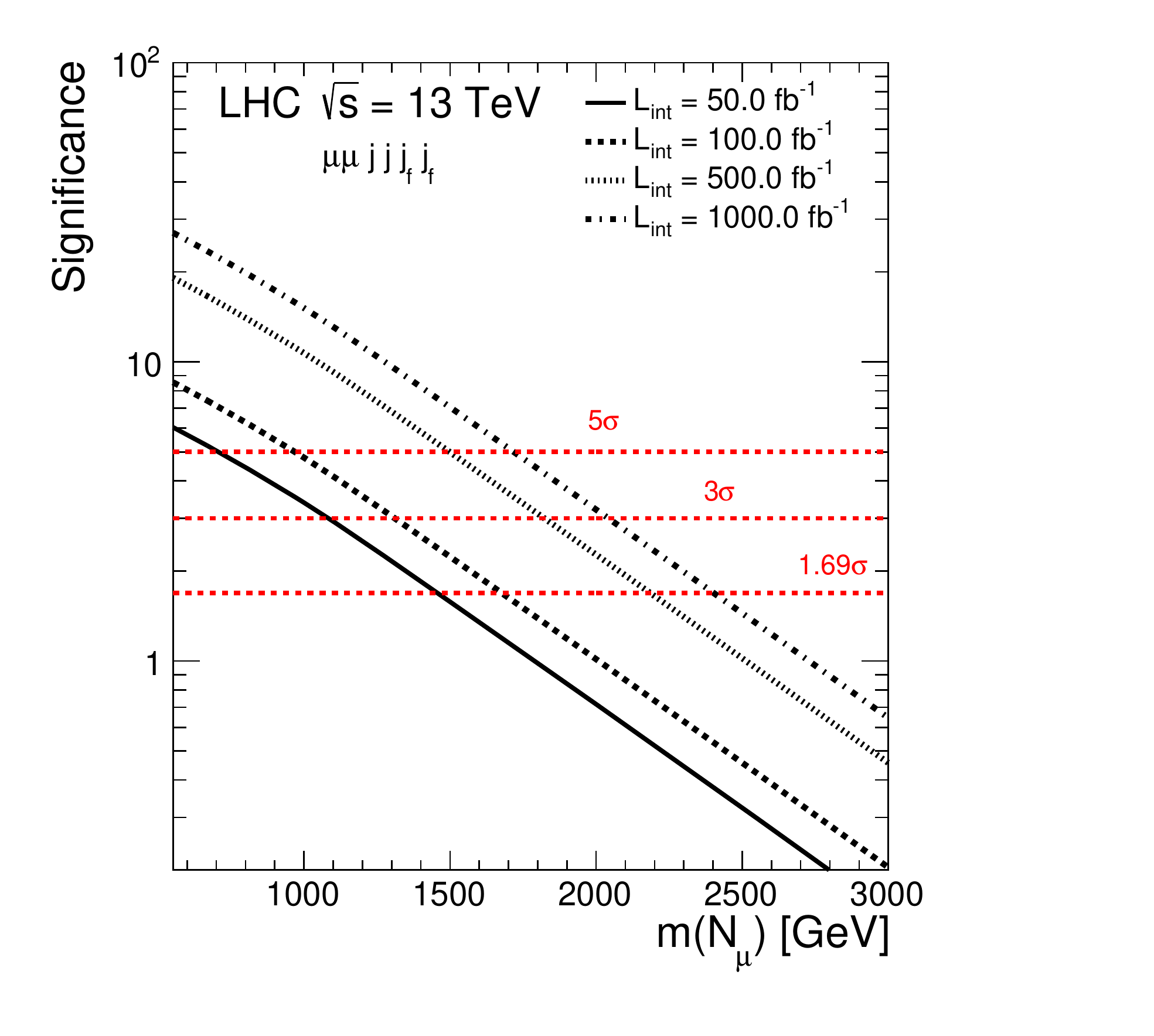}
 \end{center}
 \caption{Signal significance as a function of $m_{N_{\mu}}$ and $L_{int}$ for the $\mu\mu j j j_{f} j_{f}$ search channel.}
 \label{fig:MuMuSignificance}
 \end{figure}

\begin{figure}
 \begin{center} 
 \includegraphics[width=0.5\textwidth, height=0.35\textheight]{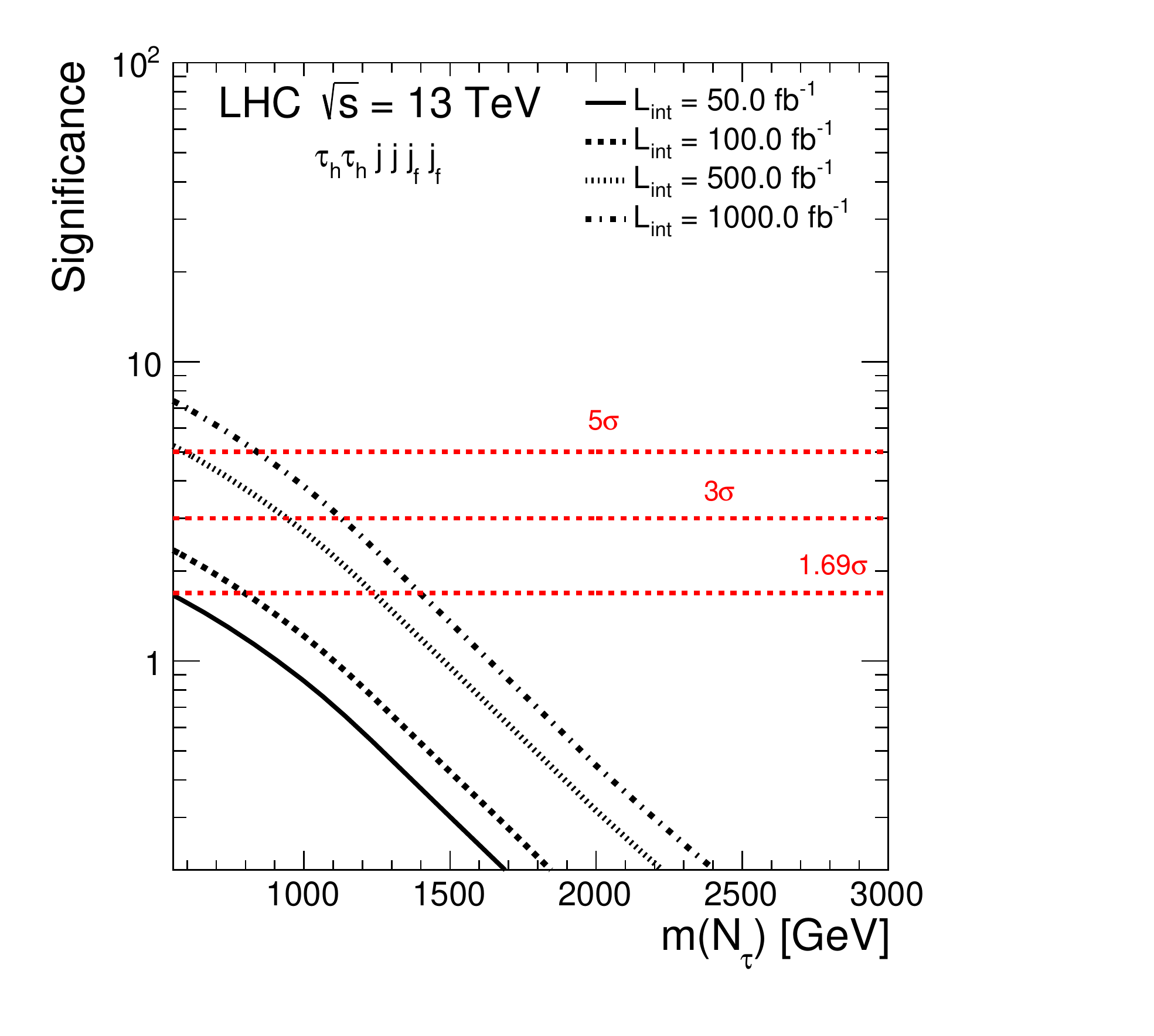}
 \end{center}
 \caption{Signal significance as a function of $m_{N_{\tau}}$ and $L_{int}$ for the $\tau_{h}\tau_{h} j j j_{f} j_{f}$ search channel.}
 \label{fig:TauTauSignificance}
 \end{figure}

Figures~\ref{fig:MuMuSignificance} and~\ref{fig:TauTauSignificance} show the expected signal significance in the $\mu\mu j j j_{f} j_{f}$ and $\tau_{h}\tau_{h} j j j_{f} j_{f}$ channels, respectively, for integrated luminosities $L_{int} = 50$-$1000$ fb$^{-1}$ and mixing $|V_{\ell N_{\ell}}|^{2}=1$. 
For $L_{int} = 100$ fb$^{-1}$, the expected 95\% exclusion on $m_{N_{\mu}}$ using the $\mu\mu j j j_{f} j_{f}$ search channel is $\sim 1.7$ TeV, while the $3\sigma$ ($5\sigma$) reach is $\sim 1.3$ (1.0) TeV. Similarly, the expected 95\% exclusion on $m_{N_{\tau}}$ using the $\tau_{h}\tau_{h} j j j_{f} j_{f}$ search channel is $\sim 0.8$ TeV at $L_{int} = 100$ fb$^{-1}$. For the longer-term projections of the high-luminosity LHC using $L_{int} = 1000$ fb$^{-1}$, the expected 95\% exclusion on $m_{N_{\mu}}$ ($m_{N_{\tau}}$) using the $\mu\mu j j j_{f} j_{f}$ ($\tau_{h}\tau_{h} j j j_{f} j_{f}$) search channel is 2.4 (1.45) TeV, while the 3$\sigma$ reach is $\sim 2.05$ (1.15) TeV. To highlight the importance of a $N_{\ell}$ search using the VBF topology, it is noted that the most recent search for $N_{\tau}$~\cite{Khachatryan:2016jqo} places no bound on $m_{N_{\ell}}$ beyond the LEP limit ($m_{N_{\ell}} < 0.1$ TeV). As noted previously, this is because the current searches (non-VBF) for $N_{\ell}$ at ATLAS and CMS are optimized for resonant $W_{R}^{\pm}$ production from DY processes ($pp\to W_{R}^{\pm} \to \ell N_{\ell}$) and assumes the  mass is accessible at the LHC (i.e. the $W_{R}$ mass is not too heavy). Existing results of the mTISM searches~\cite{ATLAS:2012ak,Khachatryan:2014dka}, which are only performed in the electron and muon channels, exclude $m_{N_{\ell}} < 500$ GeV for $|V_{\ell N_{\ell}}|^{2} = 1$. Therefore, a search for $N_{\ell}$ using VBF could be a complementary piece to the current $N_{\ell}$ search program at ATLAS and CMS, presenting a possible and important avenue for discovery.

\begin{figure}
 \begin{center} 
 \includegraphics[width=0.5\textwidth, height=0.35\textheight]{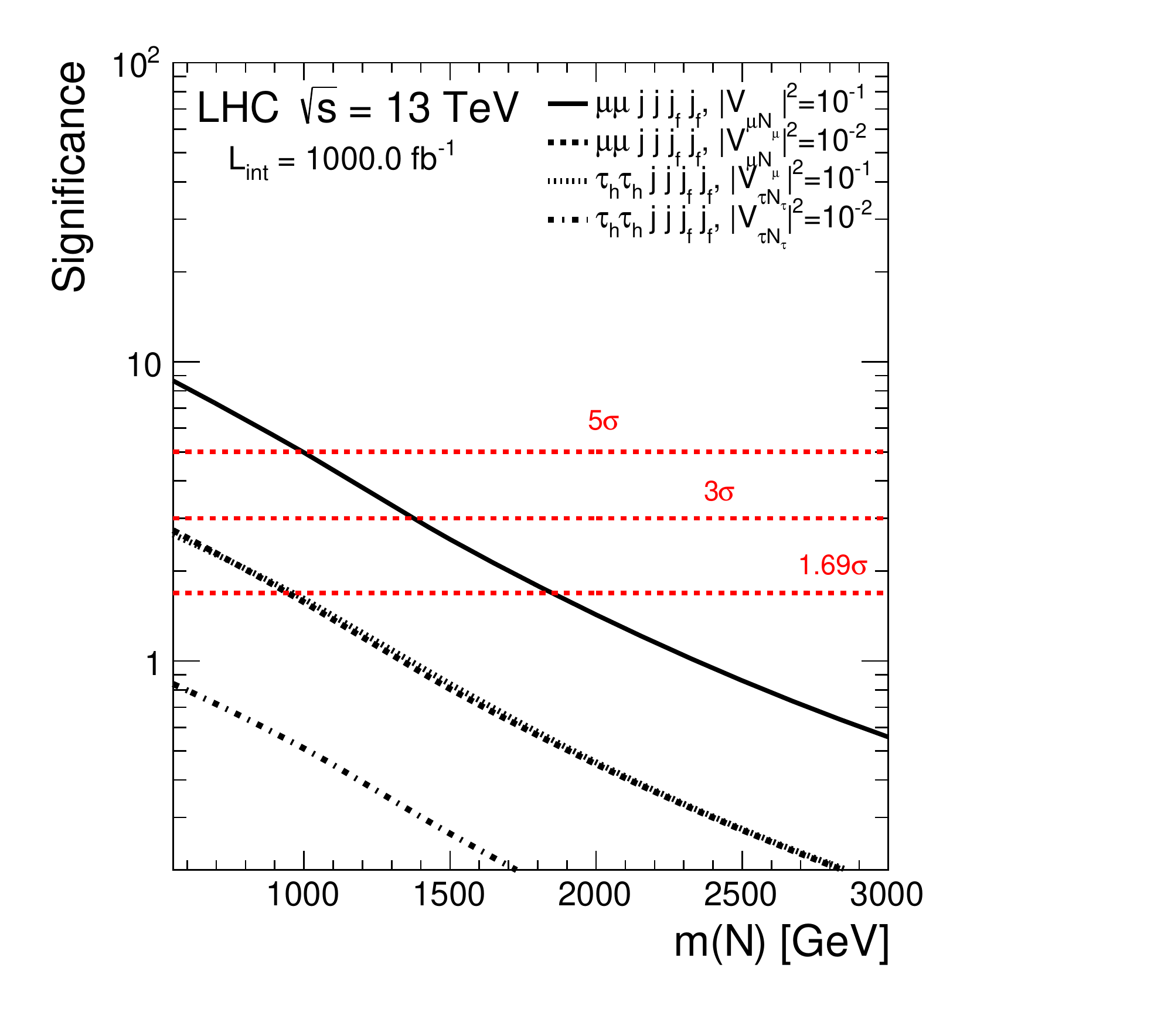}
 \end{center}
 \caption{Signal significance as a function of $m_{N_{\ell}}$ and mixing $|V_{\ell N_{\ell}}|^{2}$, assuming $L_{int} = 1000$ fb$^{-1}$.}
 \label{fig:SignificanceVsMixing}
 \end{figure}

We have optimized the selections/strategy outlined in this paper assuming the mixing between $N_{\ell}$ and $\nu_{\ell}$ is unity ($|V_{\ell N_{\ell}}|^{2} = 1$). However, our calculations can be generalized to scenarios with different mixing by appropriately scaling the expected signal yields. The event rate for $N_{\ell}$ detection at the LHC scales as $|V_{\ell N_{\ell}}|^{2}$, and so smaller mixing leads to less sensitivity. Figure~\ref{fig:SignificanceVsMixing} shows the expected signal significance in the $\mu\mu j j j_{f} j_{f}$ and $\tau_{h}\tau_{h} j j j_{f} j_{f}$ channels, assuming an integrated luminosity of $L_{int} = 1000$ fb$^{-1}$, considering different mixing scenarios. For $L_{int} = 1000$ fb$^{-1}$ and $|V_{\ell N_{\ell}}|^{2}=10^{-1}$ ($|V_{\ell N_{\ell}}|^{2}=10^{-2}$), the expected 95\% exclusion on $m_{N_{\ell}}$ is about 1.85 (1.0) TeV.

\begin{figure}
 \begin{center} 
 \includegraphics[width=0.5\textwidth, height=0.35\textheight]{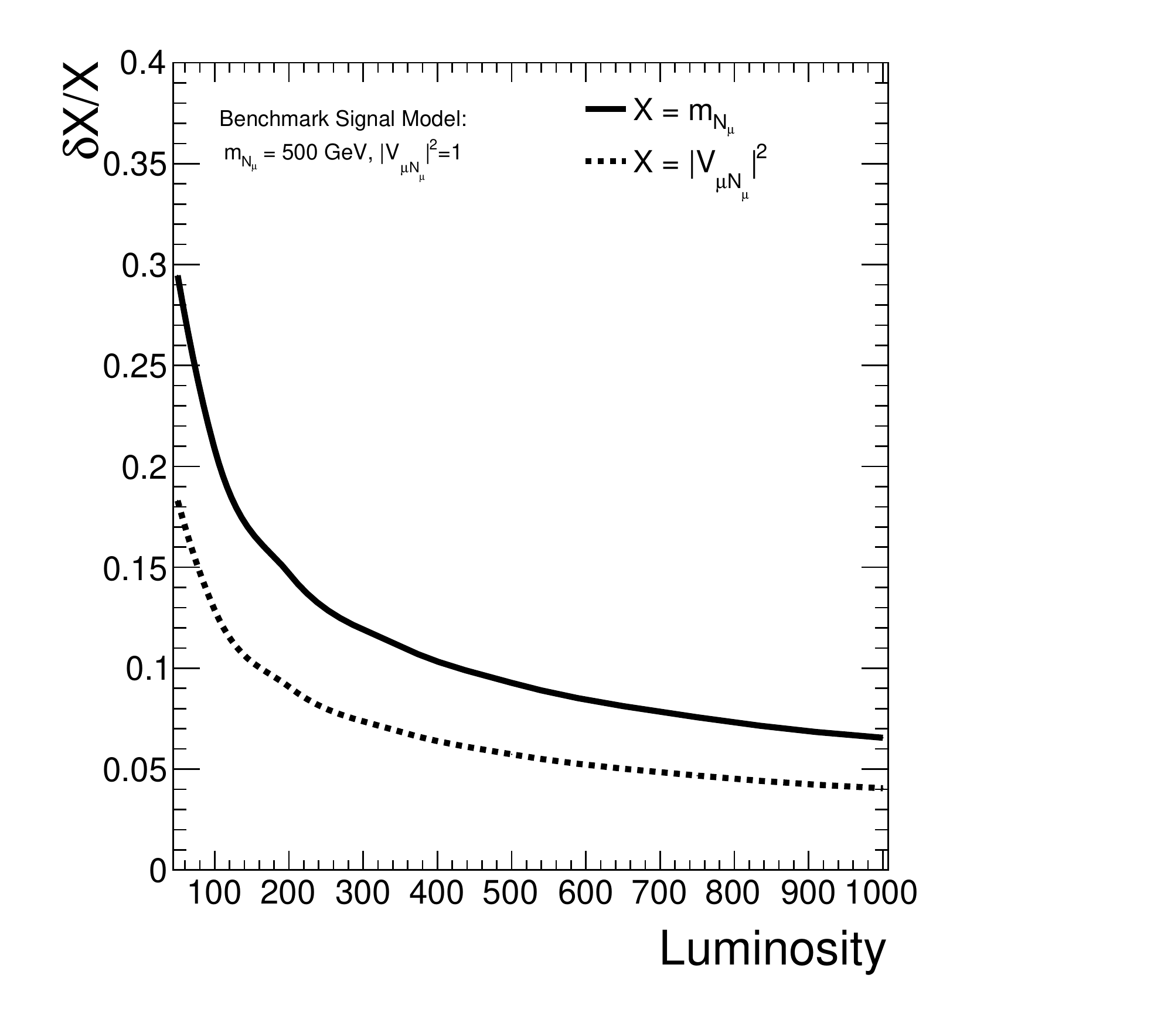}
 \end{center}
 \caption{Uncertainty on the measured heavy neutrino mass and mixing, as a function of integrated luminosity. The signal sample with $m_{N_{\mu}} = 500$ GeV and $|V_{\ell N_{\ell}}|^{2}=1$ is used as a benchmark.}
 \label{fig:UncertaintyA}
 \end{figure}

\begin{figure}
 \begin{center} 
 \includegraphics[width=0.5\textwidth, height=0.35\textheight]{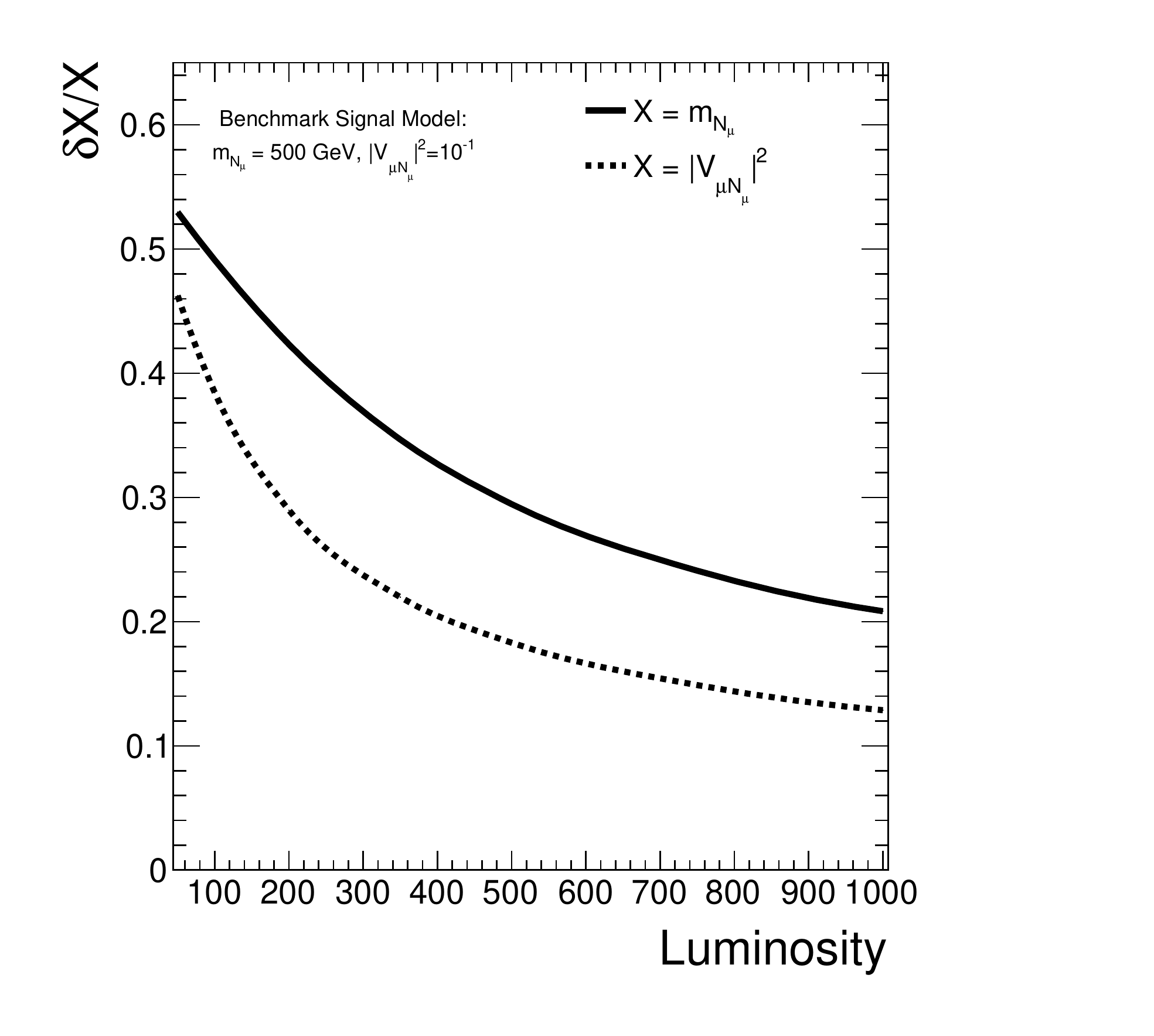}
 \end{center}
 \caption{Uncertainty on the measured heavy neutrino mass and mixing, as a function of integrated luminosity. The signal sample with $m_{N_{\mu}} = 500$ GeV and $|V_{\ell N_{\ell}}|^{2}=10^{-1}$ is used as a benchmark.}
 \label{fig:UncertaintyB}
 \end{figure}

\begin{figure}
 \begin{center} 
 \includegraphics[width=0.5\textwidth, height=0.35\textheight]{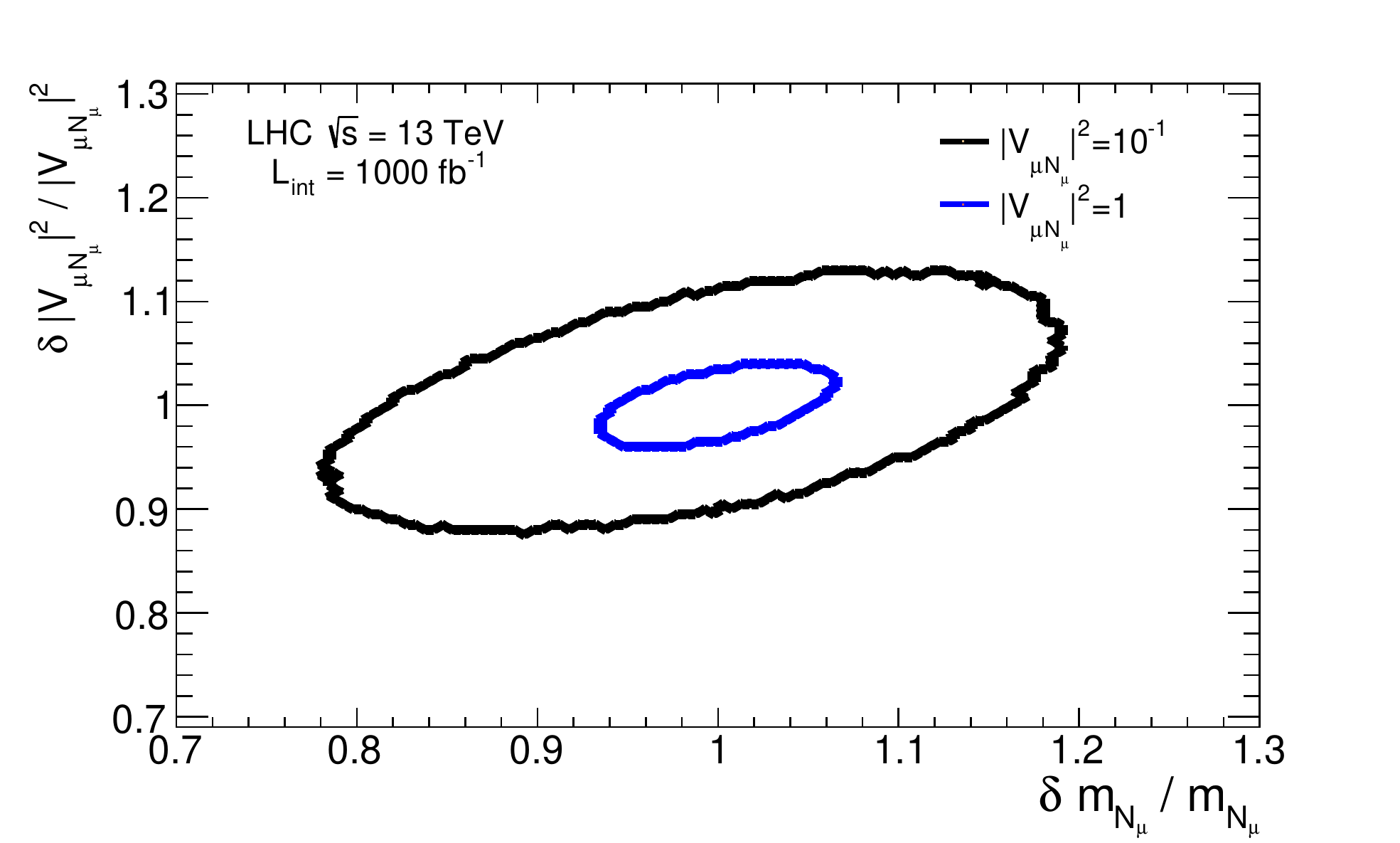}
 \end{center}
 \caption{1$\sigma$ contour lines in the $m_{N_{\mu}}$-$|V_{\mu N_{\mu}}|^{2}$ plane for the signal benchmark scenario with $m_{N_{\mu}} = 500$ GeV and $L_{int} = 1000$ fb$^{-1}$.}
 \label{fig:UncertaintyC}
 \end{figure}

From Figures~\ref{fig:STmetMuon} and~\ref{fig:STmetTau}, it is clear that varying the rate and the shape of the $S^{MET}_{T}$ distribution can be used to solve for the mass of $N_{\ell}$ as well as the mixing $|V_{\ell N_{\ell}}|^{2}$. The VBF $N_{\ell}$ study described in this paper was performed over a grid of input points in the $m_{N_{\ell}}$-$|V_{\ell N_{\ell}}|^{2}$ plane, and the $S^{MET}_{T}$ shape and expected rate in the search region were used to extract the measured $m_{N_{\ell}}$ and $|V_{\ell N_{\ell}}|^{2}$ values, including the expected uncertainties on those measurements. Figures~\ref{fig:UncertaintyA} and~\ref{fig:UncertaintyB} show how well $m_{N_{\ell}}$ and $|V_{\ell N_{\ell}}|^{2}$ can be measured as a function of integrated luminosity. For the signal benchmark scenario with $m_{N_{\mu}} = 500$ GeV and $|V_{\mu N_{\mu}}|^{2}=1$, Figure~\ref{fig:UncertaintyA} shows that the heavy neutrino mass can be determined to within 30\% (7\%) accuracy, assuming $L_{int} = 50$ (1000) fb$^{-1}$. Similarly, the mixing can be determined to within 18\% (4\%) accuracy for an integrated luminosity of $L_{int} = 50$ (1000) fb$^{-1}$. In Figure~\ref{fig:UncertaintyB} a benchmark signal sample with a lower mixing value of $|V_{\mu N_{\mu}}|^{2}=10^{-1}$ is considered. In Figure~\ref{fig:UncertaintyC}, the benchmark cases of $|V_{\mu N_{\mu}}|^{2}=1$ and $|V_{\mu N_{\mu}}|^{2}=10^{-1}$ were used to plot 1$\sigma$ contours in the $m_{N_{\mu}}$-$|V_{\mu N_{\mu}}|^{2}$ plane for an integrated luminosity of 1000 fb$^{-1}$ at the 13 TeV LHC.

\section{Discussion}

The main result of this paper is that searching for heavier neutrino states $N_{\ell}$ (predicted by extensions of the SM and inferred by the observation of neutrino oscillations) produced through VBF processes, can be a key methodology to discover $N_{\ell}$ particles at the LHC. To highlight the expanded discovery reach, we consider $N_{\ell}$ production via VBF in the context of the mTISM and focus on the $\mu\mu j j j_{f} j_{f}$ and $\tau_{h}\tau_{h} j j j_{f} j_{f}$ channels to show that the requirement of a same-sign dilepton pair combined with two central jets and two additional high $p_{T}$ forward jets with large separation in pseudorapidity and with large dijet mass is effective in reducing SM backgrounds. Assuming 100 fb$^{-1}$ of 13 TeV proton-proton data from the LHC and mixing $|V_{\ell N_{\ell}}|^{2}=1$, the expected exclusion bounds (at 95\% confidence level) are $m_{N_{\mu}} < 1.7$ TeV and $m_{N_{\tau}} < 0.8$ TeV in the $\mu\mu j j j_{f}j_{f}$ and $\tau_{h}\tau_{h} j j j_{f}j_{f}$ channels, respectively. These expected exclusion bounds with the VBF topology are to be compared to the current bounds of $m_{N_{\mu}} < 0.5$ TeV and $m_{N_{\tau}} < 0.1$ TeV. The use of the VBF topology to search for $N_{\ell}$ particles with TeV scale masses increases the discovery reach at the high-luminosity LHC, with signal significances greater than 5$\sigma$  (3$\sigma$) for $N_{\ell}$ masses up to 1.7 (2.05) TeV and 0.9 (1.15) TeV in the $\mu\mu_{h} j j j_{f}j_{f}$ and $\tau_{h}\tau_{h} j j j_{f}j_{f}$ channels, assuming $L_{int} = 1000$ fb$^{-1}$. 
It has been shown that broad enhancements in the $S_{T}^{MET}$ distributions after VBF selections provide a smoking gun signature for VBF production of $N_{\ell}$. By simultaneously fitting the $S_{T}^{MET}$ shape and observed rate in data, the mass and mixing can be measured to within 7\% and 4\% accuracy for an integrated luminosity of 1000 fb$^{-1}$ at the 13 TeV LHC.

\section{Acknowledgements}

We thank the constant and enduring financial support received for this project from the faculty of science at Universidad de los Andes (Bogot\'a, Colombia), the faculty of science at Universidad de Antioquia, the administrative department of science, technology and innovation of Colombia (COLCIENCIAS), the Physics \& Astronomy department at Vanderbilt University and the US National Science Foundation. This work is supported in part by NSF Award PHY-1506406.

\end{document}